# Optical signatures of antiferromagnetic correlations in a strongly interacting quantum Hall $MoSe_2$ monolayer

Jiho Sung[1,2,3*], Pavel A. Volkov[1,4*], Ilya Esterlis[1,5], Jue Wang[1,2], Luke N. Holtzman[6], Takashi Taniguchi[7], Kenji Watanabe[8], Katayun Barmak[6], James Hone[9], Mikhail D. Lukin[1], Philip Kim[1] & Hongkun Park[1,2†]

[1]Department of Physics, Harvard University, Cambridge, MA 02138, USA

[2]Department of Chemistry and Chemical Biology, Harvard University, Cambridge, MA 02138, USA

[3]Center for Functional Nanomaterials, Brookhaven National Laboratory, Upton, NY 11973, USA

[4]Department of Physics, University of Connecticut, Storrs, CT 06269, USA

[5]Department of Physics, University of Wisconsin-Madison, Madison, WI 53706, USA

[6]Department of Applied Physics & Applied Mathematics, Columbia University, New York, NY 10027, USA

[7]Research Center for Materials Nanoarchitectonics, National Institute for Materials Science, 1-1 Namiki, Tsukuba 305-0044, Japan

[8]Research Center for Electronic and Optical Materials, National Institute for Materials Science, 1-1 Namiki, Tsukuba 305-0044, Japan

[9]Department of Mechanical Engineering, Columbia University, New York, NY 10027, USA

*These authors contributed equally to this work

†To whom correspondence should be addressed: hpark@g.harvard.edu

**Abstract**

Strong magnetic fields quench the kinetic energy of electrons, leading to the formation of flat energy bands, known as Landau levels (LLs). In this situation, even weak interactions can drive the emergence of various ordered phases. The simplest of such phases is a quantum Hall ferromagnet, where a spontaneous spin polarization emerges when LLs with opposite spins cross. The presence of strong electron-electron interaction at zero field changes this picture and makes the resulting states much harder to predict. Here we use magneto-optical spectroscopy to reveal quantum Hall states with unconventional correlations favouring an unpolarized state in the strongly correlated electron liquid in a $MoSe_2$ monolayer. The oscillations of the exciton polaron energies as a function of perpendicular magnetic field and electron density demonstrate the emergence of LLs in a correlated electron liquid and density-dependent crossings between LLs of opposite valleys. On lowering the LL filling factor, where interactions within LLs are stronger, the crossings systematically broaden, indicating an increase in the Zeeman energy required to fully polarize the valley-degenerate LLs. These observations are shown to be consistent with antiferromagnetic interactions between LL electrons, favouring a ground state with zero valley polarization, and are therefore inconsistent with conventional quantum Hall ferromagnetism. This discovery demonstrates a qualitatively distinct form of quantum Hall magnetism in a strongly correlated electron liquid, establishing an anchoring point for understanding spin-unpolarized fractional and ordered states of correlated electrons driven by magnetic field.

## Introduction

In two-dimensional electron systems subjected to a perpendicular magnetic field, Landau quantization suppresses kinetic energy and makes Coulomb interactions the leading energy scale within each Landau level (LL). As such, LLs provide a fertile ground for interaction-driven phases with spontaneously broken symmetries. When distinct LLs are brought to degeneracy by tuning magnetic field and carrier density[1-3], the exchange interactions lift the degeneracy by the formation of quantum Hall ferromagnetic (QHFM) phases[4], which have been observed in graphene and GaAs quantum wells[5-9]. In the limit of weak electron-electron interactions where $r_s$, the dimensionless ratio between the Coulomb and kinetic energy at zero field, is much less than the unity[10,11], spontaneously spin-polarized Ising phases have been predicted when spin is the only degree of freedom at degeneracy point [4,12], consistent with experiments on GaAs quantum wells.

In the opposite limit of $r_s >> 1$, both the existence of LLs and the character of their interaction-induced phases have not been established. Two-dimensional (2D) systems are especially enigmatic in this respect. While Fermi liquid theory provides a well-established starting point for metallic states in a three-dimensional system (3D)[13,14], this theory in the low density limit is insufficient to describe the electronic behaviour in a 2D system even under weak magnetic fields[15]. On increasing $r_s$ even further, the Fermi liquid eventually gives way to an electronic (Wigner) crystal phase, where no LLs is expected to form[16-18].

Transition metal dichalcogenide (TMD) monolayers provide an ideal platform to probe the evolution of quantum Hall states on increasing $r_s$ because they feature relatively flat energy bands, owing to the large electron effective masses. As $r_s$ increases on lowering the electron density, exchange interactions enhance the spin susceptibility and hence the Zeeman splitting in a magnetic field, making the ratio between Zeeman and cyclotron energy vary smoothly with the carrier density[17]. Consequently, by simply changing the density with an electrostatic gate under a magnetic field, the energies of spin/valley resolved LLs can be controlled accurately[19,20]. In addition, due to the Ising spin-orbit coupling, the low-energy electronic states are only two-fold degenerate in the valley quantum number (K or K') with spin locked to valley. Electrons in TMDs therefore avoid complication from multiple degeneracies (due to spin, valley, sublattice) present in graphene[7-9], where more complex QHFM arises in a larger SU(4) subspace. LLs in TMD thus provide an ideal case of a two-dimensional electron gas with controlled Ising internal degree of freedom in direct comparison to that of GaAs[5,6].

Previous studies on TMD monolayers have shown a density-tunable Zeeman effect and LL crossings[21-23]. Transport experiments have reported anti-crossing features near LL degeneracies, which has been interpreted by some as signatures of quantum Hall ferromagnetism[20] while others argued such an interpretation being unlikely due to large filling factors $\nu > 20$ [19]. More recently, local compressibility measurements revealed anomalous energy gaps between LLs and hysteretic transitions consistent with QHFM at $\nu \leq 10$ [24]. These measurements have been sensitive primarily to LL gaps and thus provided indirect access to level reordering, and, moreover do not allow to resolve subtle transitions in the compressible, partially polarized regime.

Magneto-optical spectroscopy, which more directly resolves the spin/valley character of LLs, has reported abrupt changes in spin/valley polarization, but such spectral signatures were resolved only at a single LL crossing density[25]. Quantum oscillations of exciton polaron properties were also reported[26] and later attributed to nonequilibrium valley relaxation processes[27]. However, a full accounting of exchange- and correlation-driven LL realignment, as well as the detailed understanding of quantum Hall magnetic transitions and their dependence on the LL filling, has remained elusive.

Here we use magneto-optical spectroscopy on strongly correlated electrons in a $MoSe_2$ monolayer to reveal unconventional quantum Hall magnetism beyond the conventional ferromagnetic paradigm. Owing to significant Landau level mixing and strong correlations in the high $r_s$ limit, TMDs offer a completely different regime for quantum Hall magnetism (QHM). We probe Landau level crossings as a function of magnetic field and electron density for $10 < \nu < 30$. At all fillings, we observe that the degeneracy points depend on the filling of the topmost LLs, and that the valley ordering of occupied and empty LLs can be different. These effects directly demonstrate the impact of electron-electron interactions within those LLs, which cannot be captured via a simple renormalization of band parameters (*i.e.*, renormalized effective mass and $g$-factor). Secondly, through the analysis of energy shifts of attractive polaron resonances, we reconstruct Landau level filling in the K and K' valleys, revealing that the interactions within valley-degenerate Landau level favour spin-unpolarized states (Fig. 1a), as opposed to the more conventional fully polarized ferromagnetic state. We present a simple theoretical model that captures these experimental observations and enables the quantification of the intra-LL interaction strengths. Our work demonstrates that strong electronic correlations can lead to new forms of magnetism in the quantum Hall regime.

## Results

Figure 1b shows a schematic of the device, which consists of a $MoSe_2$ monolayer encapsulated between hexagonal boron nitride (hBN) layers, with top and bottom few-layer graphene gates to control the electron density. To reduce the contact resistance, an additional hexagonal boron nitride layer and a metal contact gate are added to heavily dope the $MoSe_2$ monolayer region adjacent to the bottom Pt electrode. The devices are mounted in a custom-built scanning confocal microscope based on a dilution refrigerator, which enables circular polarization-resolved optical measurements across a range of magnetic fields ($H \leq 9$ T).

In a $MoSe_2$ monolayer, the optical valley selection rule gives rise to right (σ+) and left (σ−) circularly polarized exciton-polaron resonances in the K and K' valleys, respectively (Fig. 1c). Owing to spin-valley locking arising from strong spin-orbit coupling, a positive magnetic field lifts the degeneracy at the band minima, leading to the doped electrons being spin and valley polarized (spin-down in the K' valley) at densities below a field-dependent characteristic electron density, $n_c$. In this regime, the optical response in the K' valley (σ− spectra, Fig. 1d) exhibits a weak blue shift because the electron in an exciton is indistinguishable from the doped electrons. For the K valley response, the optical spectrum (σ+ spectra, Fig. 1e) splits into higher energy repulsive polaron (RP) and lower energy attractive polaron (AP) branches because the electron in an exciton is distinguishable from doped electrons and intervalley exciton-electron interaction is dominated by polaronic effects[17,28,29].

Above $n_c$, electrons begin to occupy the K valley as well as the K' valley, entering a partially polarized regime, and low energy AP branches appear in both σ− and σ+ spectra. Concurrently, pronounced magnetic oscillations in AP branches are observed. To characterize the magnetic oscillations, we fit the AP branches of both σ− and σ+ spectra using Lorentzian functions to extract the resonance energies $E_{AP}^{K\prime}$ and $E_{AP}^{K}$ (Fig. 2c, d). The extracted AP energies, shown for a representative field of 9 T in Figs. 2c and 2d, exhibit alternating blueshifts and redshifts with increasing electron density for both σ− (K') and σ+ (K) spectra. Notably, the oscillations in the σ− and σ+ AP branches share the same periodicity, but the energy shifts are out of phase: when the σ− AP resonance energy blueshifts, the σ+ AP resonance energy shows a corresponding redshift, and vice versa. The phase relation between the σ− and σ+ AP branches is captured clearly by the sum and difference of their resonance energies, shown in the inset of Fig. 2c. Oscillatory behavior can be resolved at magnetic fields as low as 2.5 T (Extended Data Fig. 1).

We note that the observation of these features required ultrahigh-quality flux-grown $MoSe_2$ crystals[30]. We perform our measurements at a lattice temperature of 30 mK in a dilution refrigerator with excitation light power below 1 nW (Extended Data Fig. 2 for light power dependence). Similar quantum oscillations are observed in a second $MoSe_2$ monolayer device, **D2** (Extended Data Fig. 3), confirming the reproducibility of the phenomenon.

In contrast to the conventional Shubnikov-de Haas (SdH) oscillations, the magneto-oscillations we observe appear only in the partially spin-polarized regime, i.e., when both spin up (K) and spin down (K') LLs are partially occupied. This behavior can be explained by considering the effects of Landau level filling within each valley on the attractive polaron. We can approximate the AP energy in the K and K' valley at low densities[29] as $E_{AP}^{K} = \alpha n_K + \beta n_{K'}$ and $E_{AP}^{K'} = \alpha n_{K'} + \beta n_K$, where $n_K$ and $n_{K'}$ are the electron densities in the K and K' valleys respectively. Coefficients $\alpha > 0$ and $\beta < 0$ correspond to the blueshift in the AP energy due to Pauli blocking within the same valley and the redshift due to polaronic dressing by opposite valley electrons, respectively.

The oscillations in this case arise from the alternating spin (and valley) filling, as schematically explained in Fig. 2b, where the three horizontal dashed lines, labeled (i - iii), indicate representative Fermi levels. As the electron density increases and the Fermi level shifts from (i) to (ii), a LL in the K' valley is populated, while the K valley electron concentration remains constant. In this case, the σ+ AP resonances undergo redshifts, whereas the σ− AP branch blueshifts. In contrast, when the Fermi level changes from (ii) to (iii), the σ+ AP branch blueshifts, while the σ− AP branch redshifts. This leads to a sawtooth pattern of red- and blue- shifts as a function of density, which corresponds to K' or K valley being filled at a given density. In contrast, in the fully polarized regime (Fig. 2a), as the gate voltage increases, the electron density in K' valley increases linearly in voltage, and consequently, the σ+ AP resonance energy decreases monotonically even in the presence of sharp changes in the density of states associated with the LL formation.

Consistent with the description above, the derivative of the fitted σ+ AP resonance energy with respect to electron density, $dE_{AP}^{K}/dn$, (Fig. 2e), shows that most oscillations in the partially polarized regime are confined between two limiting values, corresponding to the coefficients $\alpha$ and $\beta$, with the negative $\beta$ close to the slope observed in the fully polarized regime. The sign of $dE_{AP}^{K}/dn$ thus enable the determination of which valley is being occupied at a given density.

However, one feature in Fig. 2e cannot be explained by this simple picture: a pronounced enhancement near $n \sim 3.4 \times 10^{12}$ cm$^{-2}$ (red circle). This feature is related to the sudden blueshift of $E_{AP}^{K}$ together with sudden redshift of $E_{AP}^{K\prime}$, as marked in Fig. 2c-d. As we will explain below, these features are associated with the nearly degenerate LLs in the K and K' valleys.

To elucidate the origin of the abrupt blueshift, we performed σ+ resolved reflection measurements as a function of electron density and magnetic field. In Fig. 3a, we show the map of the extracted $dE_{AP}^{K}/dn$ that reveals characteristic Landau fan-like features[19,20,24]. In the remainder of the text, we use $dE_{AP}/dn$ as a shorthand for $dE_{AP}^{K}/dn$. The boundaries between regions of positive and negative $dE_{AP}/dn$ coincide with the filling of integer number $\nu$ of LLs, as indicated by green dotted lines showing $n = \nu \frac{B}{h/e}$, where $h$ and $e$ denote Planck's constant and electron charge. The period of the observed magneto-oscillations establishes the presence of Landau quantization at $r_s$ up to approximately ~10.

The observation of Landau quantization in our system should not be taken for granted. The magnetic field values we use results in the underlying $r_s$ value is 5 ~ 10 and a large LL mixing parameter, $\kappa = E_{\text{Coulomb}}/E_{\text{cycl}} = m_e^* e/(4\pi\varepsilon_0 \varepsilon l_B \hbar B) = \alpha/\sqrt{H[\text{Tesla}]}$, which is the ratio of the Coulomb energy to the cyclotron gap. Here, $m_e^*$, $\varepsilon_0$, $\varepsilon$, $l_B$, and $\hbar$ denote the effective mass of electrons, vacuum permittivity, dielectric constant, magnetic length (with $l_B = \sqrt{\hbar/eB}$) and reduced Planck constant, respectively. $\alpha$ is a material dependent constant and is 56.5 for a hBN-encapsulated $MoSe_2$ monolayer, obtained from the quantum Monte Carlo (QMC) approach discussed below. The large $\kappa$ and $r_s$ values indicate that the LLs in our system come from a strongly correlated electronic liquid, and their properties may thus differ from LLs formed at small $r_s$. This is in contrast to the LLs formed in more conventional 2DEGs, such as GaAs heterostructures or graphene, where $\kappa$ and $r_s$ are typically of order unity[31,32], which differs from the strongly correlated TMD systems we study in this work.

The map of $dE_{AP}/dn$ in Fig. 3d, which is a magnified view of the range enclosed by the yellow box in Fig. 3a, shows enhanced responses as a function of electron density and magnetic field. These enhancements are confined to the density range marked by the black arrow. Outside this density range, $dE_{AP}/dn$ shows a diagonal stripe pattern with alternating bright ($dE_{AP}/dn > 0$, filling K valley) and dark ($dE_{AP}/dn < 0$, filling K' valley) stripes. This behavior is reproduced schematically in Fig. 3e, which shows the valley state map of the LL at the Fermi energy (K valley: blue; K' valley: red), obtained from the sign of the $dE_{AP}/dn$ response. When $n$ crosses ~$3.4 \times 10^{12}$ cm$^{-2}$, the stripe contrast flips. Equivalently, when one follows a given diagonal stripe (*e.g.,* within filling factor, $\nu_{tot}$ = 17 and 18) on increasing density, its contrast changes, indicating a change in the relative ordering of the K and K' LLs, as illustrated in Fig. 3b and c.

The enhanced $dE_{AP}/dn$ can therefore be interpreted as a signature of degeneracy points for the K and K' valley LLs. In this case, even small changes of experimental parameters result in the population transfer of the topmost filled LL electrons from the K' to K valley (Fig. 3f). We note that we only observe enhanced positive $dE_{AP}/dn$, corresponding to transfer from K' to K, and not from K to K'. This observation implies that, with increasing $n$, the energies of the K valley LLs always decrease relative to the those in the K', leading to an increase in the population at K.

The observed pattern of switching and enhanced responses can be attributed to the density dependence of the spin susceptibility and hence the Zeeman energy. As the density increases, the ratio of the Zeeman energy $\Delta_Z$ to the cyclotron energy $\hbar\omega_c$ decreases due to the decreasing impact of Coulomb interactions (Fig. 3b, c). In Fig. 3g, we plot the effective Zeeman splitting between the topmost occupied LL at each valley, $\Delta_Z^{eff}$, color coded as a function of *H* and *n*, obtained from the QMC spin susceptibility (see SI Sec. 1 & 2). The K' and K valley LLs switch their relative ordering with increasing density and cross at a vertical line of constant electron density, where $\Delta_Z^{eff} = 0$ (indicated by the arrow and black dashed line). The density at which this LL degeneracy occurs in the QMC calculations is close to the experimentally observed value, ~$3.4 \times 10^{12}$ cm$^{-2}$ in the magnetic field range 7-9 T.

There is, however, one crucial difference between the experimental and QMC results. In the experiment, the strong $dE_{AP}/dn$ responses appear as tilted line segments that extend over a finite range of densities, as opposed to a vertical *n* = *constant* line in the QMC result. By comparing the experimental valley state map (Fig. 3e) with the QMC predictions (Fig. 3h), we can see that the experimental valley states differ from QMC over certain ranges of density and magnetic field (enclosed by black dotted lines in Fig. 3h); we refer to these as the "valley-switching regions". For example, a particularly sharp discrepancy between Fig. 3e and 3h occurs at points along the $\nu_{tot}$ = 19 line (a representative point marked by red circle). In the QMC prediction, this point corresponds to a crossing where the stripes switch their valley states, whereas the experimental data indicate that the Fermi energy remains in the K' valley LL all around that point (Fig. 3e). Starting from this point, both increasing and decreasing the magnetic field lead to changes in the K' valley population (Fig. 3i), indicating that both the lowest-energy unoccupied states and highest-energy occupied states are in the K' valley.

The explanation of the experimental results therefore requires the following considerations beyond the renormalized band picture: (1) filling-dependent LL degeneracy position to explain the tilting of $dE_{AP}/dn$ maxima away from $n$ = *constant* line (2) the reversal of the filled and empty LL ordering near degeneracy (Fig. 3i). This strong filling dependence of the LL ordering cannot be captured in a renormalized band picture and requires treatment of intra-LL interactions for $r_s >> 1$ and strong LL mixing. We note that a proper account of such effects remains an open challenge[15].

To explain the evolution of the observed LL filling, we introduce a phenomenological model that captures the main details of experiment. Specifically, we consider the energy of the two highest partially occupied LLs, with $N_{K'}$ and $N_K$ electrons in K' and K valleys, respectively. The interaction energy can be modeled as $\frac{1}{2N}\left( U_{K'}N_{K'}^2 + U_K N_K^2 + 2\,U_{K,K'}N_K N_{K'}\right)$, where $U_{K'}$ ($U_K$) and $U_{K,K'}$ describe the intra- and inter-valley interaction energy, respectively. Introducing the valley imbalance $\delta = (N_{K'} - N_K)/N$ and an effective filling factor $\nu_{eff} = (N_{K'} + N_K)/N$, defined modulo the fully occupied LLs ($0 \leq \nu_{eff} \leq 2$), the energy per electron including the effective Zeeman term (the energy offset between K and K' LL energies) takes the form:

$$\frac{E_{tot}}{N} = \frac{U_0}{2}{\nu_{eff}}^2 + \frac{X}{2}\delta^2 + \left(\frac{U_{K'} - U_K}{4}(\nu_{eff} - 1) + \frac{\Delta_Z^{eff}}{2}\right)\delta \qquad \text{Eq. (1)}$$

where $U_0 = \frac{U_{K\prime}+U_K+2U_{K,K\prime}}{8}$ and $X = \frac{U_{K\prime}+U_K-2U_{K,K\prime}}{4}$ are the charge and spin/valley interactions, respectively, while $\frac{U_{K\prime}-U_K}{4}(\nu_{eff}-1)$ describes the exchange energy difference between the crossing LLs due to unequal LL index. In Fig. 4a, we present $d\nu_{eff,K}/d\nu_{eff}$, obtained by minimizing Eq. (1) at a fixed $\nu_{eff}$ and $\Delta_Z^{eff}$. Bright (dark) regions correspond to filling of the K (K') valley. This representation enables a direct comparison (see SI Sec. 5.2) with the experimentally determined $dE_{AP}/dn$ near ~ 3.4 × $10^{12}$ $cm^{-2}$, plotted as a function of $\nu_{tot}$ and $\Delta_Z^{eff}$ in Fig. 4b. The model reproduces both the enhanced $dE_{AP}/dn$ signal near $\Delta_Z^{eff} = 0$ and the valley-switching regions in ($\Delta_Z^{eff}, \nu_{tot}$) space (see Extended Fig. 4). The bright lines of enhanced $dE_{AP}/dn$ (experimental LL degeneracy lines) at various filling factors form linear features in ($\Delta_Z^{eff}, \nu_{tot}$) space, from which we obtain their slopes, $d\Delta_Z^{eff}/d\nu_{tot}$.

In Fig. 4c, we plot the experimentally obtained $d\Delta_Z^{eff}/d\nu_{tot}$ where the enhanced $dE_{AP}/dn$ occurs at the three LL crossing densities (indicated by the three black arrows in Fig. 3a), corresponding to 2.5, 3.4, and 5.4×$10^{12}$ $cm^{-2}$, respectively (see Fig. 4b and Extended Fig. 5). We also estimate the value of magnetic fields corresponding to the enhanced $dE_{AP}/dn$ (*e.g.,* ~ 8 T for the green circle in Fig. 3d). In the picture of LL crossing, these magnetic field values correspond to $\nu_{eff} = 1$, which we write $H_{\nu eff=1}$. Fig. 4c displays experimentally obtained $H_{\nu eff=1}$ versus $\left|d\Delta_Z^{eff}/d\nu_{tot}\right|$ for three different densities above. Within the fixed density, the extracted values are linearly proportional to the magnetic field. Moreover, when plotted as a function the total valley (and spin) polarization, $\zeta = (n_{K'}-n_K)/(n_{K'}+n_K)$, all points collapse onto a single linear dependence (Fig. 4c, inset, see SI for obtaining ζ). In our theoretical model above, the LL degeneracy is reached when $\frac{U_{K'}-U_K}{4}\left(\nu_{eff}-1\right)+\frac{\Delta_Z^{eff}}{2}=0$, namely when the coefficient of the valley imbalance $\delta$ in Eq. (1) vanishes. Therefore, the experimentally obtained slope $d\Delta_Z^{eff}/d\nu_{tot}$ is equal to $-(U_{K'}-U_K)/2$ . We note that the Hartree-Fock predictions for $U_{K'}-U_K$ both reproduce the scaling behavior in the inset of Fig. 4c and give $(U_{K'}-U_K)/2$ in the correct order of magnitude compared with the experimentally obtained $d\Delta_Z^{eff}/d\nu_{tot}$ (1.4 meV from Hartree-Fock at $H_{\nu eff=1}$ = 8 T and $n$ ~ 3.4 × $10^{12}$ $cm^{-2}$, compared with ~ 0.47 meV in Fig. 4c; see SI Sec. 5.3).

We now discuss the possibility of a magnetic transition occurring at a LL degeneracy. At weak LL mixing, spin-degenerate LLs are expected to undergo a ferromagnetic transition[12] (corresponding to $X<0$ in Eq. (1)). Such a transition is predicted to be first order in field and thus exhibit hysteresis, as illustrated by the model results in the top and middle panels of Fig. 4b. Experimentally, however, we do not observe any hysteresis in AP spectra under upward and downward electron density sweeps (Fig. 4d). Instead, the enhancement in $dE_{AP}/dn$ exhibits a finite width in $\Delta_Z^{eff}$, as shown in Fig. 4e.

Remarkably, a finite-width enhancement can occur in Eq. (1) for antiferromagnetic interactions ($X>0$), as shown in the bottom panel of Fig. 4a. Consistent with this picture, Fig. 4f shows that the width of the $dE_{AP}/dn$ peak as a function of $\Delta_Z^{eff}$ is larger at lower filling factors (and lower densities) across all three LL crossing regimes. The full width of half maxima (FWHM) of these peaks are nearly three times larger for lower density than that of higher density. Since the interaction effects (and therefore $X$) are expected to be stronger for lower density, this experimental observation is consistent with the picture for the interaction driven magnetic transition in these degenerated LLs.

We emphasize that this change in transition width cannot be attributed to disorder alone. Because LL broadening scales inversely with carrier mobility[33,34], the previous report[23] suggests that broader LLs and wider transition peaks should occur at higher densities, which is opposite to our observations. In our device, the quantum mobility obtained from Dingle analysis over the density range $2.0 \times 10^{12}$ cm$^{-2}$ < $n$ < $3.5 \times 10^{12}$ cm$^{-2}$ remains essentially constant (Extended Fig. 6). To study the interplay between disorder and interactions, we incorporate the effect of disorder phenomenologically into Eq. (1) via a random gaussian distribution of single-particle energies with width Γ[35,36] (SI Sec. 6) and present the $d\nu_{eff,K}/d\Delta_Z^{eff}$ (proportional to the measured $dE_{AP}/dn$ due to dependence of $\Delta_Z^{eff}$ on $n$, see SI Sec. 6.1) as a function of $\Delta_Z^{eff}$ in Fig. 4g. For ferromagnetic interactions ($X < 0$ in Eq. 1), the model predicts a narrowing of the transition followed by the emergence of hysteresis with further increase in interactions. In contrast, for antiferromagnetic interactions, the transition width increases for stronger interactions, which is fully consistent with our observations. This observation indicates that in our system, antiferromagnetic interactions between electrons at pseudospin-degenerate LLs lead to a valley-unpolarized ($N_K$=$N_{K'}$), thus spin unpolarized ground state. (Fig. 1a).

**Discussion and outlook**

Both the mechanism driving the antiferromagnetic interaction and the ultimate form of the quantum ground state constitute important questions. Previously, antiferromagnetic intervalley coupling has been considered in twisted graphene[37]. In that case, however, the wavefunctions can be spatially separated at two different sublattices, analogous to quantum Hall bilayers, where an interlayer coherent state has been established[12].

As for the ground state, while the fully polarized state is only two-fold degenerate, there is an extensive number of field-unpolarized candidate states for any $0 < \nu_{eff} < 2$. As an example, (canted) anti-ferromagnetic QHFM has been observed in both monolayer and bilayer graphene [38-40]. However, while the spin-configuration can be anti-ferromagnetically aligned, this graphene based QHFM should be viewed as SU(4) ferromagnetism, where the uniform ordering occurs in the combined spin-valley SU(4) flavor space. The electronic wavefunctions of $MoSe_2$ reside on a single sublattice only, analogous to the usual spin-degenerate LL problem known to have a ferromagnetic ground state[12] (SI Sec. 5.3). Another unpolarized candidate state is an intervalley coherent state, where all electrons would be in a uniform superposition between two spin/valley states, similar to a QHFM state with an in-plane magnetization[12]. We find no evidence for translational symmetry breaking and associated umklapp peaks[16,17] that would occur in intervalley coherent states (at wavevector K-K') or charge density wave states at partial LL fillings. Finally, while antiferromagnetism has been shown to allow for quantum anomalous Hall transport (even in the absence of Landau quantizing external fields)[41,42], our results demonstrate, in contrast, that strong magnetic fields can induce antiferromagnetism in a correlated system.

What distinguishes $MoSe_2$ from other systems is the higher effective mass, leading to the large $r_s$ regime and strong LL mixing where Coulomb interaction dominates. This consideration suggests that the unconventional pseudospin magnetism we observe in this work can be connected to the magnetism of the proximate Wigner crystal phase[16,17,43]. Near the quantum melting transition, the crystal is known to be strongly magnetically frustrated[44] due to competing ferromagnetic and antiferromagnetic couplings. Such frustration may be connected to the antiferromagnetic coupling observed here.

Our experimental data and theoretical analyses indicate the emergence of antiferromagnetic interactions in a system with strong Landau level mixing and strong correlations at zero field, establishing it as an integral part of quantum Hall physics at strong coupling. Our work thus provides a clear starting point for further studies using other experimental probes, such as scanning tunneling microscopy for direct visualization in $MoSe_2$ monolayers[8,9]. Intriguingly, our observations provide evidence for antiferromagnetic LL correlations across a range of fillings, opening a possibility to explore spin-singlet fractional quantum Hall states[45,46] at higher magnetic fields or quantum Hall spin liquids[47] without the need for moiré superlattice. Extending these approaches to bilayer systems will enable the study of strong interlayer correlations and ordering phenomena associated with layer pseudospin degree of freedom[12,48]. More broadly, studies here and future systematic studies of quantum Hall magnetic transition in $MoSe_2$ monolayer will provide a foundation for understanding magnetism in moiré semiconductor flat bands[49,50].

**Methods:**

**Device fabrication and operation**

Monolayer $MoSe_2$, hBN flakes, and few-layer graphite layers were exfoliated from bulk crystals onto 285 nm $SiO_2$/Si substrates. $MoSe_2$ monolayers were identified by an optical microscope. The flakes were stacked by the dry transfer method using a polydimethylsiloxane stamp and a thin layer of polycarbonate. Pre-patterned Pt electrodes (10 nm) serving as electrical contacts to the $MoSe_2$ were deposited on top of the bottom hBN/few-layer graphite stack. The heterostructure was formed by stacking, in sequence, top hBN/few-layer graphite/middle hBN/$MoSe_2$ monolayer and then stamping this stack onto the pre-patterned Pt electrodes on the bottom hBN/few-layer graphite. Finally, electrical contacts and contact gates were deposited by electron-beam evaporation of Cr (10 nm)/Au (90 nm) to connect pre-patterned Pt electrodes and few-layer graphite gates. Thicknesses of hBN flakes in **D1** were measured by atomic force microscopy to be 62 nm and 67 nm for the bottom and middle layers, respectively. **D2**, contains only a bottom graphite gate, with bottom hBN thickness 51 nm.

To dope the monolayer $MoSe_2$ at dilution refrigerator temperatures, we grounded the contacts to $MoSe_2$, applied a voltage $V_{\mathrm{bg}}$ and $V_{\mathrm{tg}}$ to the bottom and top gates and at the same time illuminated the whole sample with a broadband light to activate the contacts. After doping was finished, the activation light was removed, and the sample was thermalized for 0.5 s. The optical measurements were performed after the thermalization was finished. The doping onset voltage $V_0$ is determined as the voltage at which the reflectance contrast spectra starts to deviate from the neutral regime. The electron density is calculated from the parallel-plate capacitor model $n = \varepsilon_0\varepsilon_r((V_{\mathrm{tg}} - V_0)/d_{\mathrm{tg}} + (V_{\mathrm{bg}} - V_0)/d_{\mathrm{bg}})/e$, where $\varepsilon_0$ is the vacuum permittivity, $\varepsilon_r$ is the dielectric constant of the hBN, and $d_{\mathrm{tg}}$ and $d_{\mathrm{bg}}$ is the thickness of the top and bottom hBN dielectric. We find that $\varepsilon_r = 3.1$ provides good agreement between AP magneto-oscillations and Landau fan-like features.

**Optical measurements**

Reflectance spectroscopy and magneto-optical measurements were performed with a home-built scanning confocal microscope based on a dilution refrigerator (Bluefors) with which the sample lattice temperature can reach 16 mK. The sample is mounted in the center of a superconducting magnet (AMI) capable of applying ±9 T perpendicular magnetic field. A piezo-electric stage (attocube) was used to precisely position the sample. The microscope consists of an apochromatic cryogenic objective (attocube LT-APO/LWD, NA 0.65), two fused silica plano-convex lenses (OptoSigma) at around 4 K and 50 K, two achromatic doublet lenses (Thorlabs) and a galvo scanner (Thorlabs). A compensated full-wave liquid crystal retarder (Thorlabs LCC1413-B) was placed in the shared light path of incoming and outcoming beams to impose the same ±λ/4 retardance to both beams without mechanical movement. A polarimeter (Thorlabs PAX1000IR1) was used to confirm the circular polarization outside of the dilution refrigerator. To ensure a polarization-independent light path inside the dilution refrigerator, we examined the reflection isotropy of a bare $SiO_2$/Si substrate mounted inside with a 360° rotation of the linearly polarized light. For reflectance spectroscopy, a tungsten-halogen lamp (Thorlabs SLS201L) filtered to 720 ~ 800 nm range was coupled to a single mode fiber, collimated with an objective (Olympus PLN 10×, NA 0.25) and directed to the sample, creating a diffraction-limited spot. The light power on the sample was variable but always kept below 0.7 nW. Power dependent measurements were performed using a single mode fiber coupled superluminescent diode. The reflected light was collected by a spectrometer with a 1200 g/mm grating and a CCD camera (Princeton Instruments BLAZE).

**Reflectance contrast spectroscopy**

Reflectance contrast is defined as $R_{\mathrm{C}}^{*}(E) = \frac{I(E)}{I_r(E)} - 1$, where $I(E)$ is the measured reflected light intensity and $I_r(E)$ is a reflected reference light spectrum. Two different reference spectra were used depending on the analysis. A spectrum taken at high electron density was used as the reference for the RP resonance, and a spectrum taken near zero electron density was used as the reference for the AP resonance. To extract the RP and AP resonance parameters, we fit the RP/AP resonance with a Lorentzian function $R_{\mathrm{C}}^{\mathrm{RP/AP}}(n, E) = \frac{A^2}{(E-E_0)^2+\gamma^2/4}\left[\frac{\gamma}{2}\cos\alpha - (E - E_0)\sin\alpha\right] + C$, where $A^2$, $E_0$ and $\gamma$ are the amplitude, energy and linewidth of the resonance.

**Data availability**

The data that support the plots within this paper and other findings of this study are available from the corresponding authors upon reasonable request.

**Acknowledgements**

We acknowledge useful discussions with A. Vishwanath, P. Nosov, G. Chen and P. E. Dolgirev. We acknowledge support from AFOSR (FA9550-21-1-0216), NSF CUA (PHY-2317134 for H.P. and M.D.L.), Samsung Electronics (for H.P. and P.K.), NSF (PHY-2012023 for H.P. and M.D.L), DOE (DE-SC0020115 for H.P. and M.D.L. and DE-SC0012260 for P.K.). J.S. acknowledges support by the Quantum materials press (QPress) of the Center for Functional Nanomaterials (CFN), which is a U.S. Department of Energy Office of Science User Facility, at Brookhaven National Laboratory under Contract No. DE-SC0012704. I.E. acknowledges support from the NSF through the University of Wisconsin Materials Research Science and Engineering Center Grant No. DMR-2309000. $MoSe_2$ was synthesized by J.H., K.B. and L.H. with the support of the Columbia University Materials Science and Engineering Research Center (MRSEC), through NSF grant no. DMR-2011738. K.W. and T.T. acknowledge support from the JSPS KAKENHI (Grant Numbers 21H05233 and 23H02052), CREST (JPMJCR24A5), JST, and World Premier International Research Center Initiative (WPI), MEXT, Japan.

**Author contributions**

J.S. and H.P. conceived the project. J.S. fabricated the samples. J.S. and J.W. designed and performed the experiments. P.A.V. and I.E. performed theoretical modelling and calculations. J.S., P.A.V, and I.E. analysed the data. T.T. and K.W. provided hBN crystals. L.N.H., K.B. and J.H. provided $MoSe_2$ crystals. J.S., P.A.V., I.E., and H.P. wrote the manuscript with extensive input from the other authors. H.P., P.K. and M.D.L. supervised the project.

**Competing financial interest**

The authors declare no competing interests.

**Additional Information**

Correspondence and requests for materials should be addressed to hpark@g.harvard.edu

## Figures

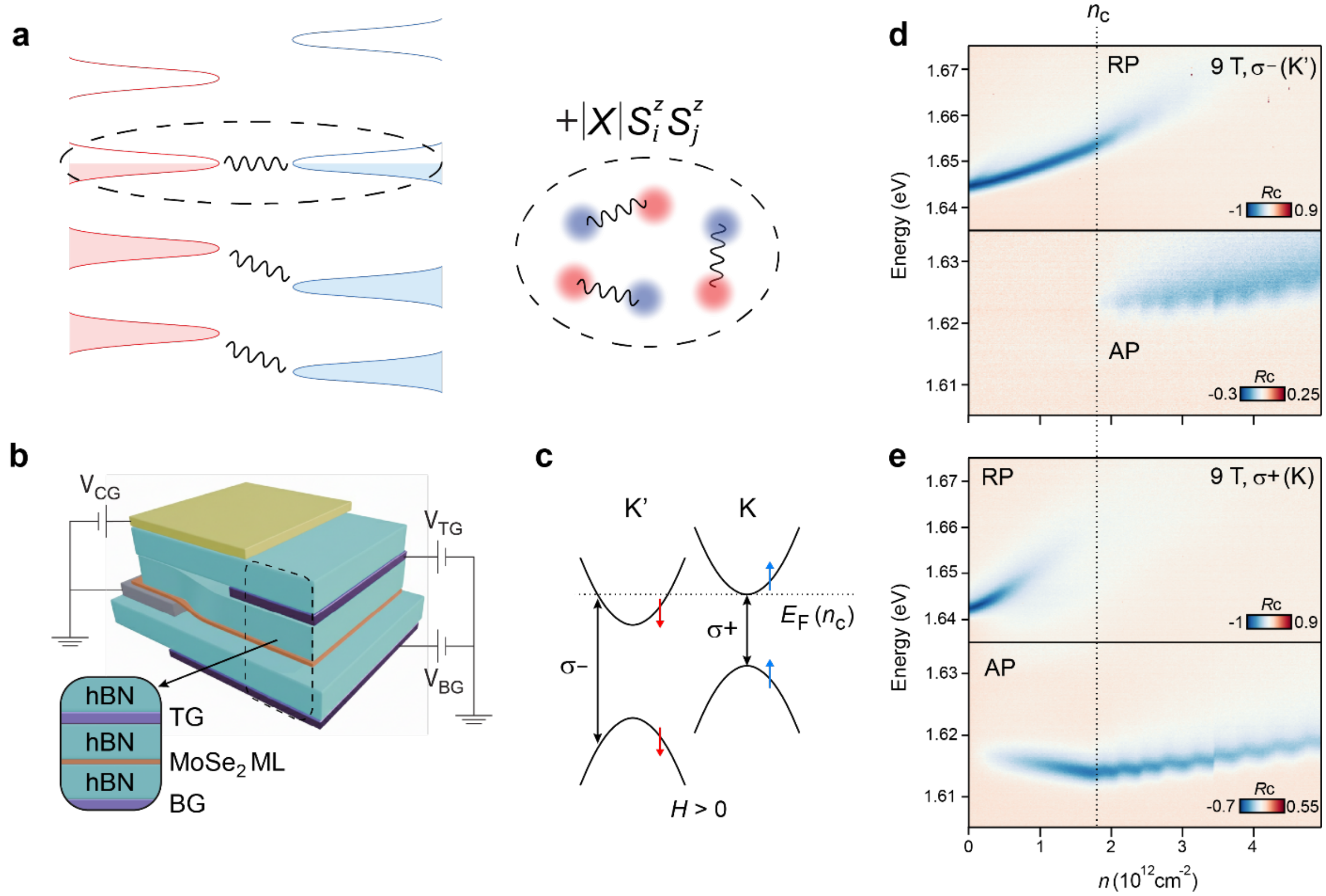

**Fig. 1 | Exciton-polaron spectroscopy of a quantum Hall liquid**

**a**, Schematic of LL diagram and exchange coupling in a strongly correlated two-dimensional electron liquid with antiferromagnetic correlations.

**b**, Schematic of a $MoSe_2$ monolayer device, consisting of a $MoSe_2$ monolayer, encapsulated by hexagonal boron nitride, with few-layer graphite serving as top and bottom gates. Platinum is used for the contact electrode to $MoSe_2$ monolayer, and additional hexagonal boron nitride and gold contact gate are added on top.

**c**, Schematic of the band edge energy alignment at the K and K' valleys under a positive magnetic field, illustrating spin-valley locking and the optical valley selection rule.

**d-e**, Left (**d**) and right (**e**) circularly polarized reflectance contrast spectra at 9 T and at a base lattice temperature of 30 mK as a function of electron density. The black dashed lines denote the characteristic density, $n_c$ above which electrons start to fill the K valley.

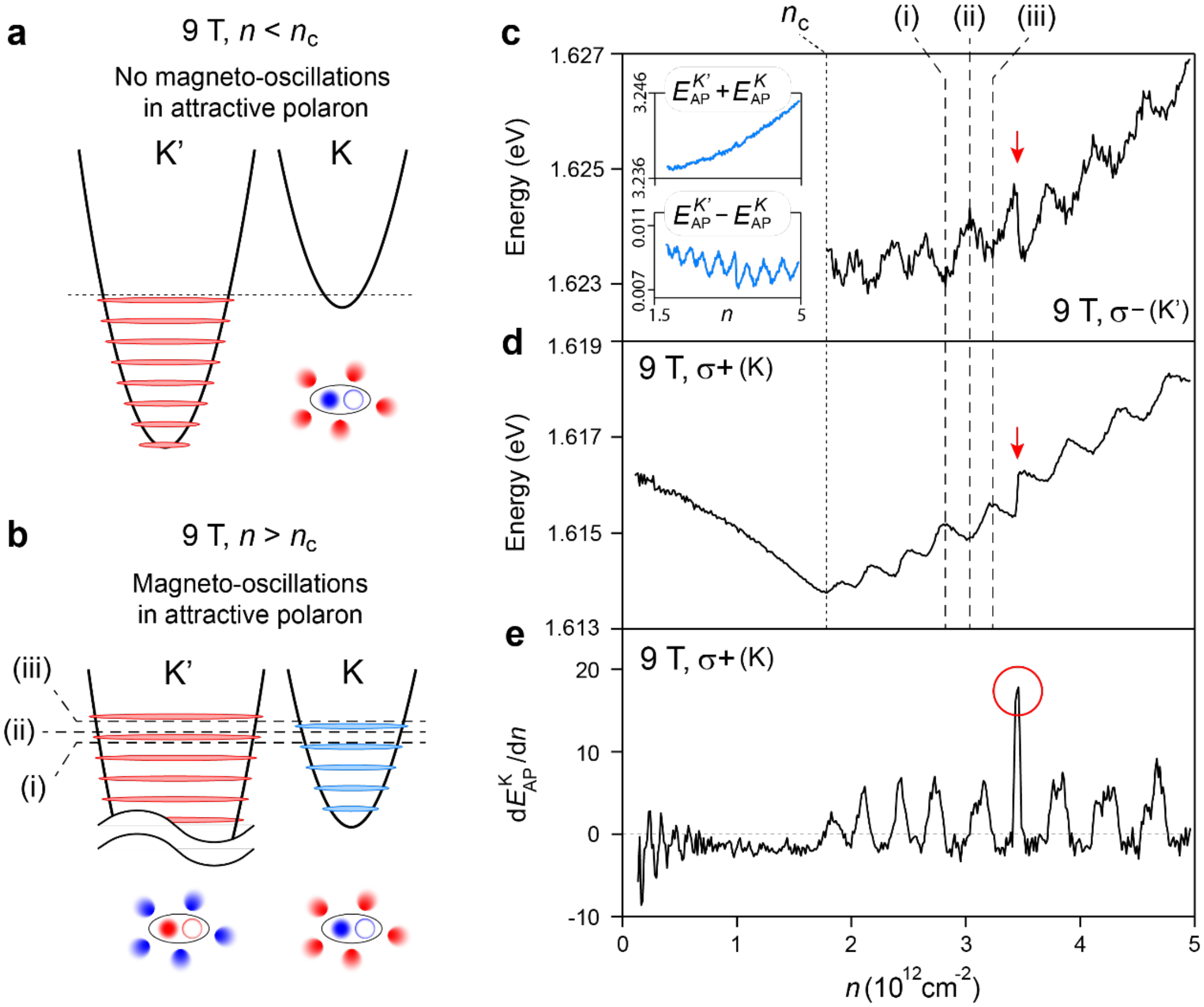

**Fig. 2 | Quantum oscillations of attractive polaron energies**

**a-b**, Schematic illustrations of the lowest conduction bands in a $MoSe_2$ monolayer at 9 T. Below $n_c$ (**a**), attractive polarons are observed only in the σ+ response and exhibit no magnetic oscillations. Above $n_c$ (**b**), attractive polarons appear in both σ− and σ+ responses, accompanied by pronounced magnetic oscillations. Fermi level positions (i), (ii), and (iii) are indicated as black dashed lines.

**c-d**, Fitted σ− (**c**) and σ+ (**d**) AP resonance energies as a function of electron density at 9 T. Black dashed lines correspond to the electron densities at which the Fermi level aligns with the positions indicated in (**b**). Insets: sum and difference of σ− and σ+ AP resonance energies as a function of electron density. Red arrows mark the abrupt redshift (**c**) and blueshift (**d**) in AP resonance energy near $n \sim 3.4\times10^{12}$ cm$^{-2}$.

**e**, $dE_{AP}/dn$ for the σ+ response as a function of electron density. The black dotted line marks zero value for $dE_{AP}/dn$. Red circle indicates the peak in $dE_{AP}/dn$ response near $n \sim 3.4\times10^{12}$ cm$^{-2}$.

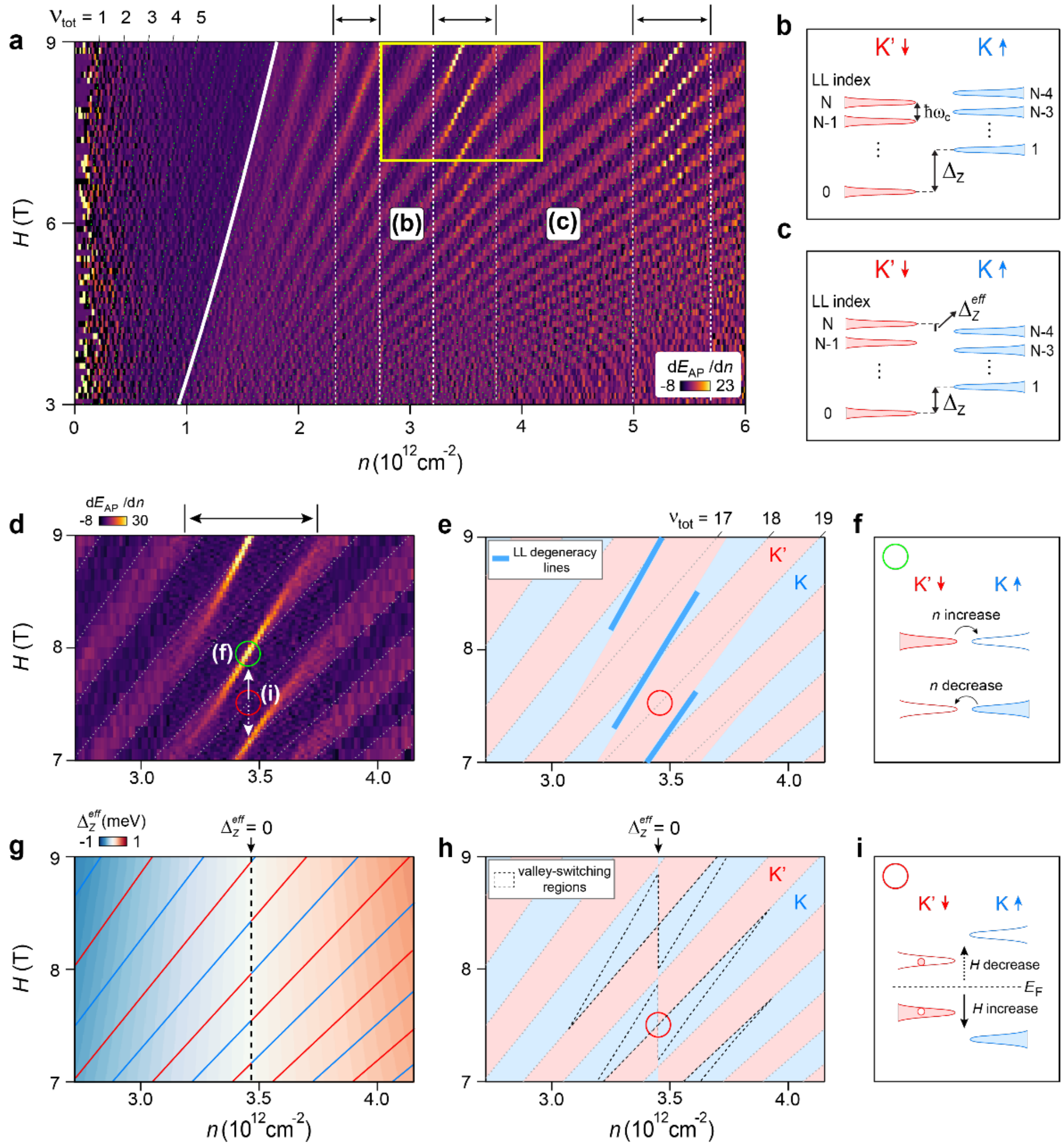

## Fig. 3 | Landau level crossings and interaction-driven realignment

**a**, Map of the derivative of the fitted σ+ AP resonance energy with respect to electron density, $dE_{AP}/dn$ as a function of electron density and perpendicular magnetic field. Green dotted lines denote the filling of integer filling $\nu_{tot}$ of LLs. Black arrows on top and white dotted lines indicate the crossing density ranges where highest partially occupied LLs at K and K' valleys become nearly degenerate.

**b-c**, Schematic LL diagrams corresponding to the regions indicated as (**b**) and (**c**) in the map. $\Delta_Z$ decreases as electron density increases. $\Delta_Z^{eff}$ is the effective Zeeman splitting between the highest occupied LLs in the K and K' valleys.

**d**, Magnified view of the $dE_{AP}/dn$ map near the LL crossing regime. Gray dotted lines indicate the filling of an integer number of LLs. Green circle indicates the center point of the LL degeneracy line, which crosses the $\nu_{tot}$ = 18 line. Red circle in (**d**, **e**, **h**) marks the point on the $\nu_{tot}$ = 19 line at the center of the degenerate density range.

**e**, Valley-state map of the LL at the Fermi energy in the same (*n, H*) space. Total filling factors are labeled above. Blue lines denote LL degeneracy lines that manifest as strong $dE_{AP}/dn$ responses in the data.

**f**, Schematic diagram of LLs at the degeneracy point marked by the green circle in (**d**), where a small increase of the density leads to a collective transfer of electrons from the K' valley LL into the K valley LL.

**g**, Colormap of $\Delta_Z^{eff}$, the effective Zeeman splitting, in the same (*n, H*) space. A black arrow and a black dashed line indicate the LL crossings along a vertical line of constant electron density. Integer filling factor lines are colored by the valley of the topmost filled LL (red for K' and blue for K).

**h**, Valley-state map of the LL in the same (*n, H*) space, predicted by QMC. Black dotted lines enclose valley switching regions, within which exchanging the valley labels reproduces the experimentally determined map in (**e**).

**i**, Schematic diagram of LLs at the point marked by the red circle in (**d**), showing an inverted ordering between occupied and unoccupied states. A solid arrow indicates the behavior under increasing magnetic field (holes occupied) and a dotted line indicates decreasing magnetic field (electrons occupied).

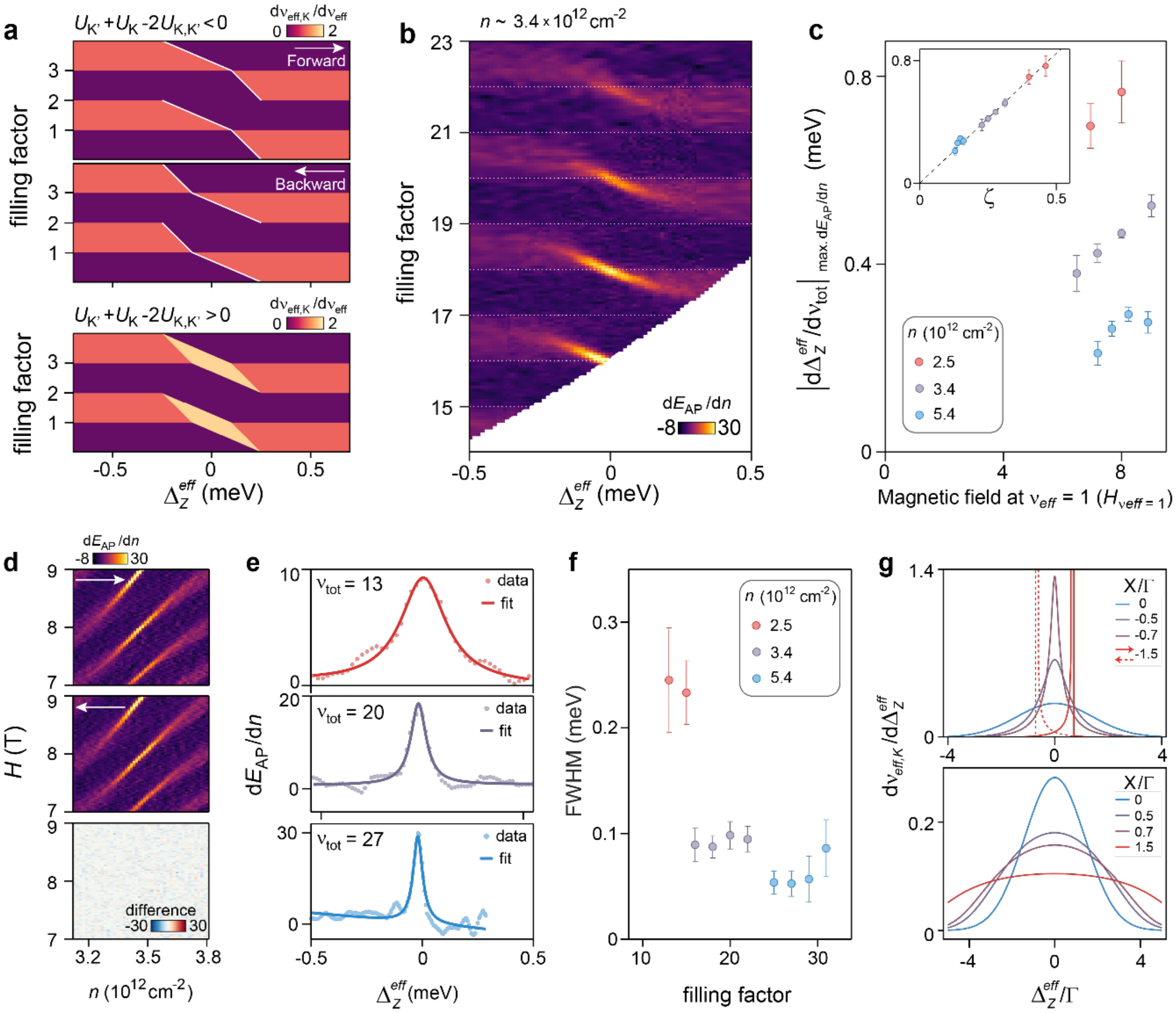

**Fig. 4 | Antiferromagnetic correlations in a strongly interacting quantum Hall liquid**

**a**, Model results for the derivative of electron occupation in the K valley, $\nu_{eff,K}$ (calculated by minimizing the total interaction energy) with respect to total occupation $\nu_{eff}$. The top and middle panels show results for $U_{K\prime} = 1$ $U_K = 0.5$, and $U_{K,K\prime} = 0.85$, corresponding to ferromagnetic interaction $\mathrm{X} = \frac{U_{K\prime}+U_K-2U_{K,K\prime}}{4} < 0$ for forward (top) and backward (middle) $\Delta_Z^{eff}$ sweeps. The bottom panel shows the antiferromagnetic (X > 0) case with $U_{K\prime} = 1$, $U_K = 0.5$, and $U_{K,K\prime} = 0.65$.

**b**, Map of the derivative of the fitted σ+ AP resonance energy with respect to the electron density, $dE_{AP}/dn$, shown versus $\Delta_Z^{eff}$ and total filling factor $\nu_{tot}$.

**c**, Extracted slopes $d\Delta_Z^{eff}/d\nu_{tot}$ obtained by fitting the enhanced $dE_{AP}/dn$ (LL degeneracy lines) in the $(\Delta_Z^{eff}, \nu_{tot})$ plane for the three LL crossing regimes. Inset: The same $d\Delta_Z^{eff}/d\nu_{tot}$ values plotted as a function of valley (and spin) polarization, $\zeta = (n_{K'}-n_K)/(n_{K'}+n_K)$.

**d**, $dE_{AP}/dn$ maps for increasing (top) and decreasing (middle) density sweeps. The difference map (bottom) between the two sweep directions shows the absence of hysteresis.

**e**, Line cuts of $dE_{AP}/dn$ along $\nu_{eff}$ = 1 for the three LL crossing regimes. Representative line profiles together with Lorentzian fits are shown for $\nu_{tot}$ = 13 near n ~ $2.5 \times 10^{12}$ $cm^{-2}$, $\nu_{tot}$ = 20 near n ~ $3.4 \times 10^{12}$ $cm^{-2}$, and $\nu_{tot}$ = 27 near n ~ $5.4 \times 10^{12}$ $cm^{-2}$.

**f**, The full-width at half-maximum (FWHM) of the Lorentzian fit, representing the transition width, as a function of total filling factor.

**g.** $d\nu_{eff,K}/d\Delta_Z^{eff}$ as a function of $\Delta_Z^{eff}$ in presence of LL broadening by disorder Γ and intra-LL exchange interaction X. The top panel corresponds to ferromagnetic interaction (X < 0) and the bottom panel to antiferromagnetic interaction (X > 0).

## Extended Data Figures

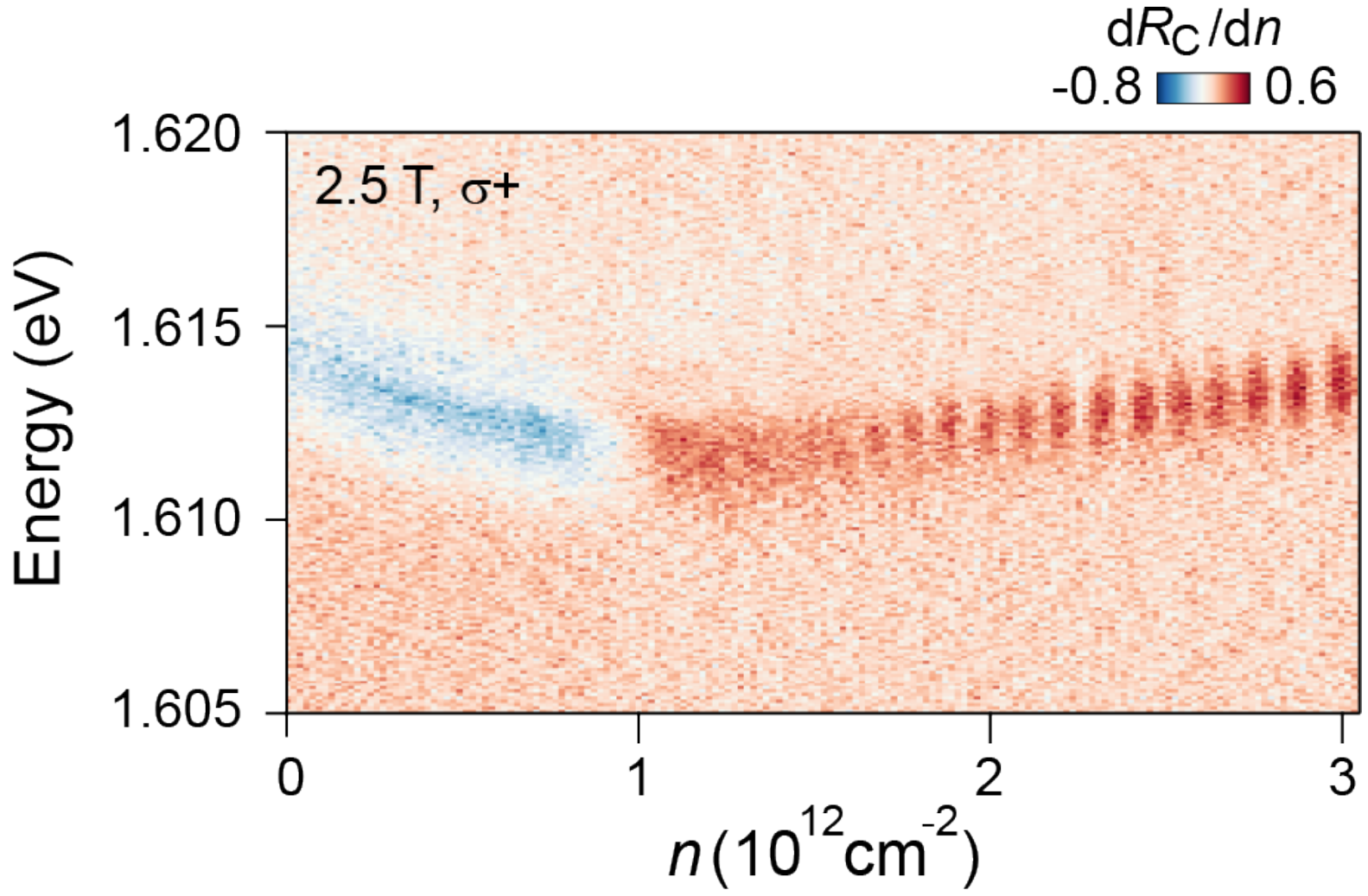

**Extended Data Fig.1 | Magneto-oscillations in the low-field regime**

Color map of the derivative of reflectance contrast spectra (2.5 T, σ+) with respect to electron density. Oscillatory behavior is observed above $n \sim 1.5\times10^{12}$ cm$^{-2}$.

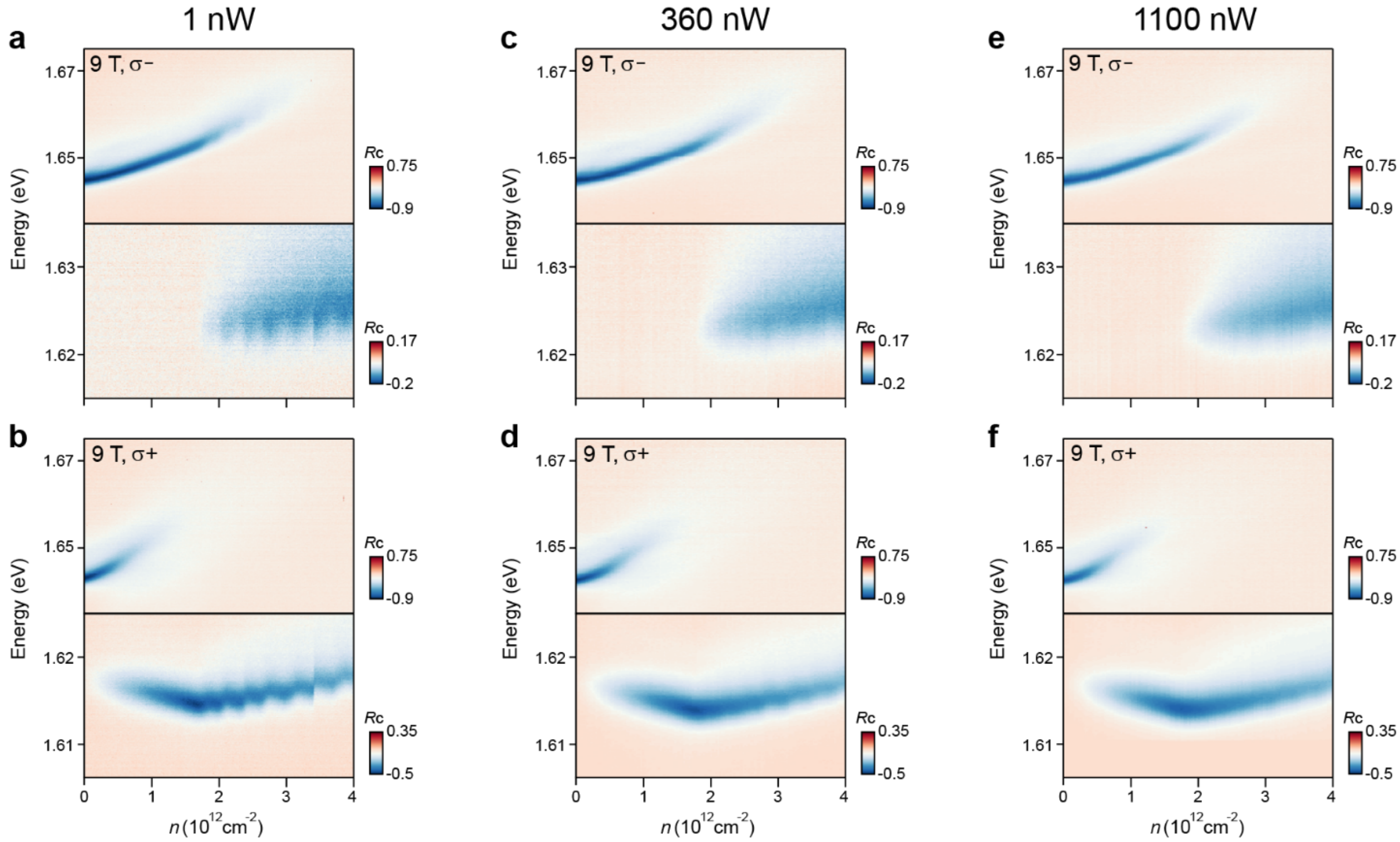

**Extended Data Fig.2 | Excitation power dependence of exciton-polaron spectra**

Left (top panels) and right (bottom panels) circularly polarized reflectance contrast spectra at 9 T and at a base lattice temperature of 30 mK as a function of electron density for different incident powers (a,b) 1 nW, (c,d) 360 nW, and (e,f) 1100 nW.

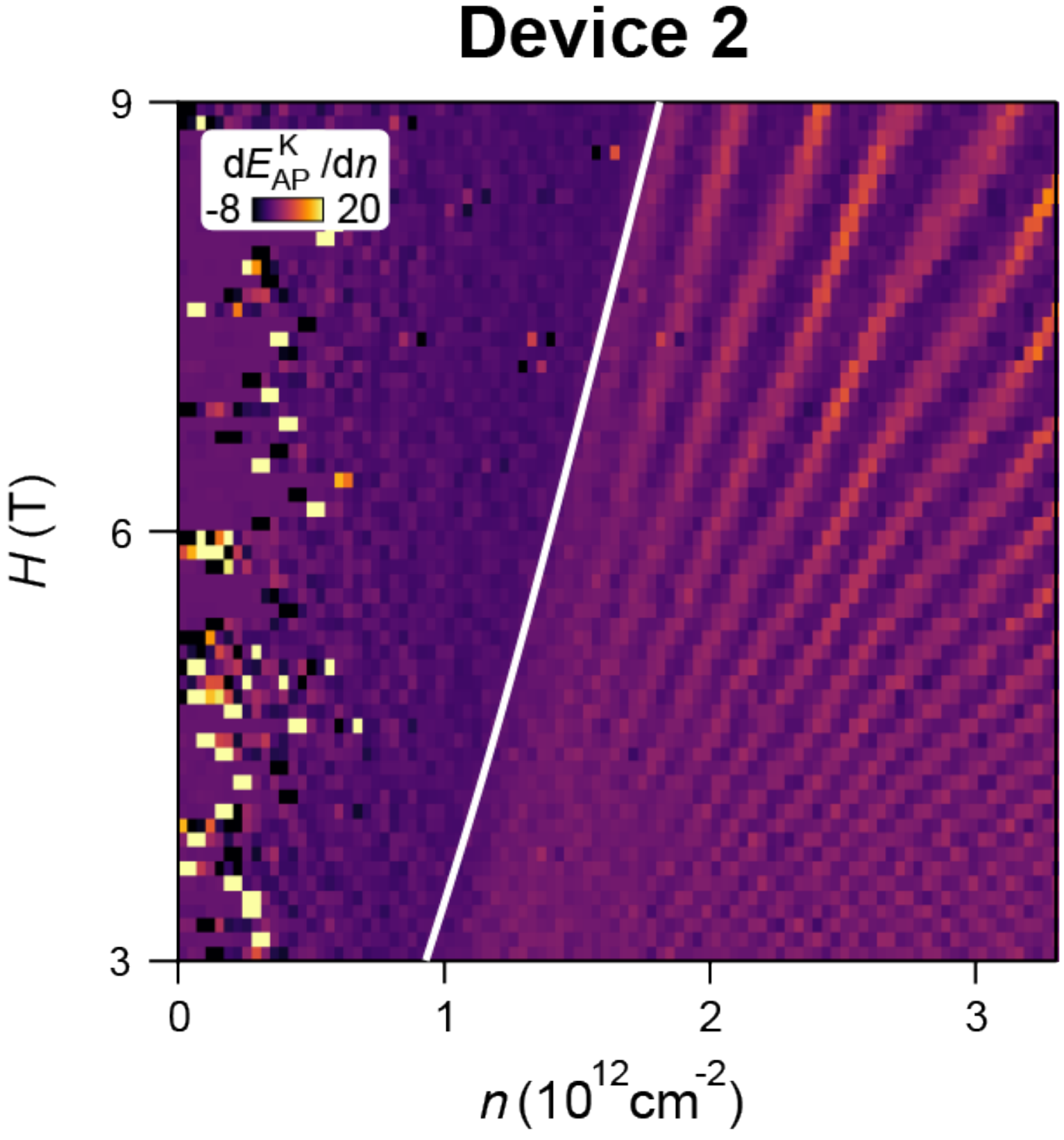

**Extended Data Fig.3 | Reproducibility of Landau fan-like features in a second device**

Map of the derivative of the fitted σ+ AP resonance energy with respect to electron density, $dE_{AP}/dn$ as a function of electron density and perpendicular magnetic field in **D2**.

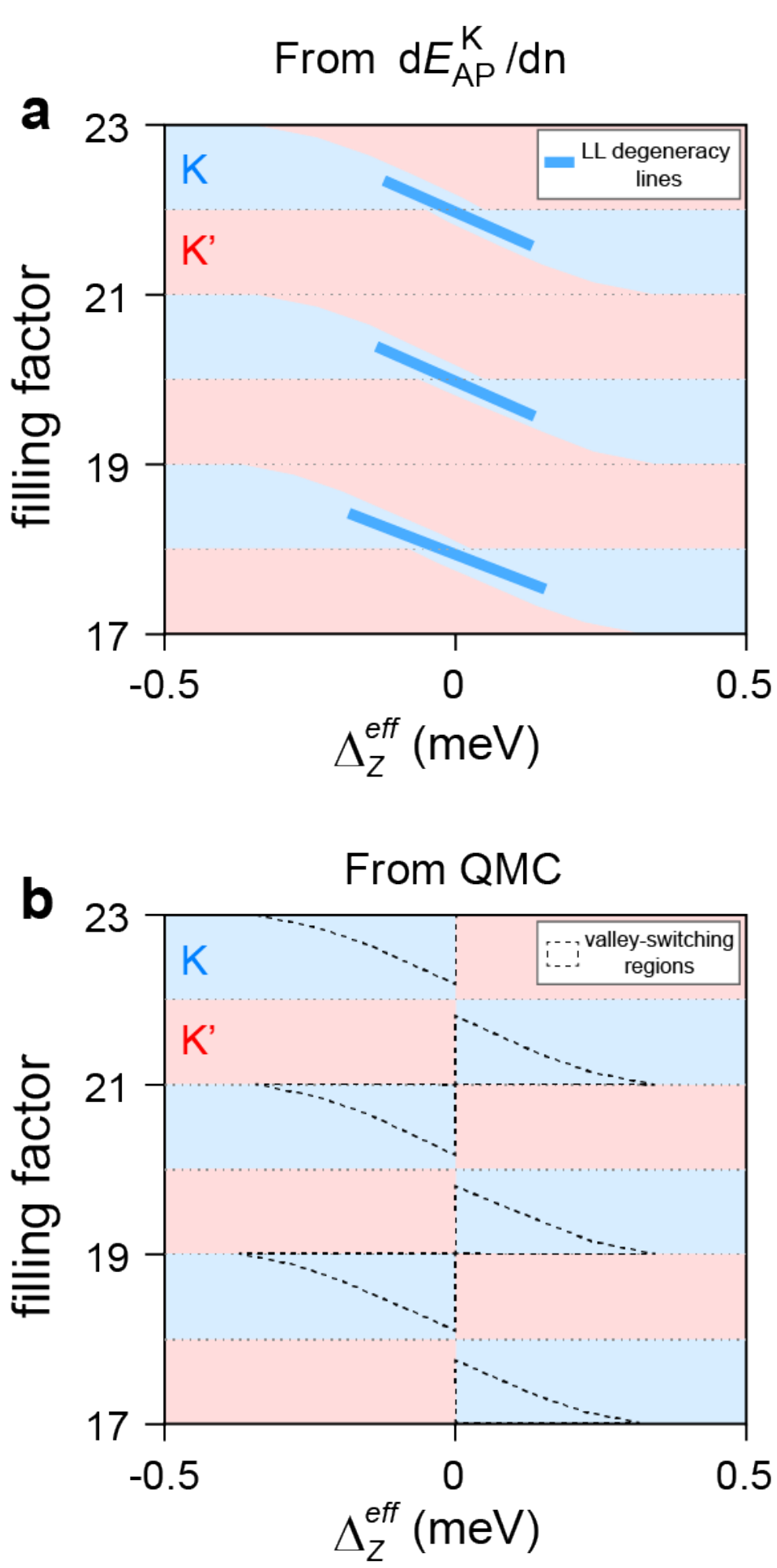

**Extended Data Fig. 4 | Comparison of experimentally determined and QMC predicted valley-state maps**

(a) Valley-state map of the LL at the Fermi energy in the $(\Delta_Z^{eff}, \nu_{tot})$ space from $dE_{AP}/dn$ data. Blue lines denote LL degeneracy lines that manifest as strong $dE_{AP}/dn$ responses in the data. (b) Valley state map of the LL in the same $(\Delta_Z^{eff}, \nu_{tot})$ space, predicted by QMC. Black dotted lines enclose valley switching regions, within which exchanging the valley labels reproduces the experimentally determined map in (a).

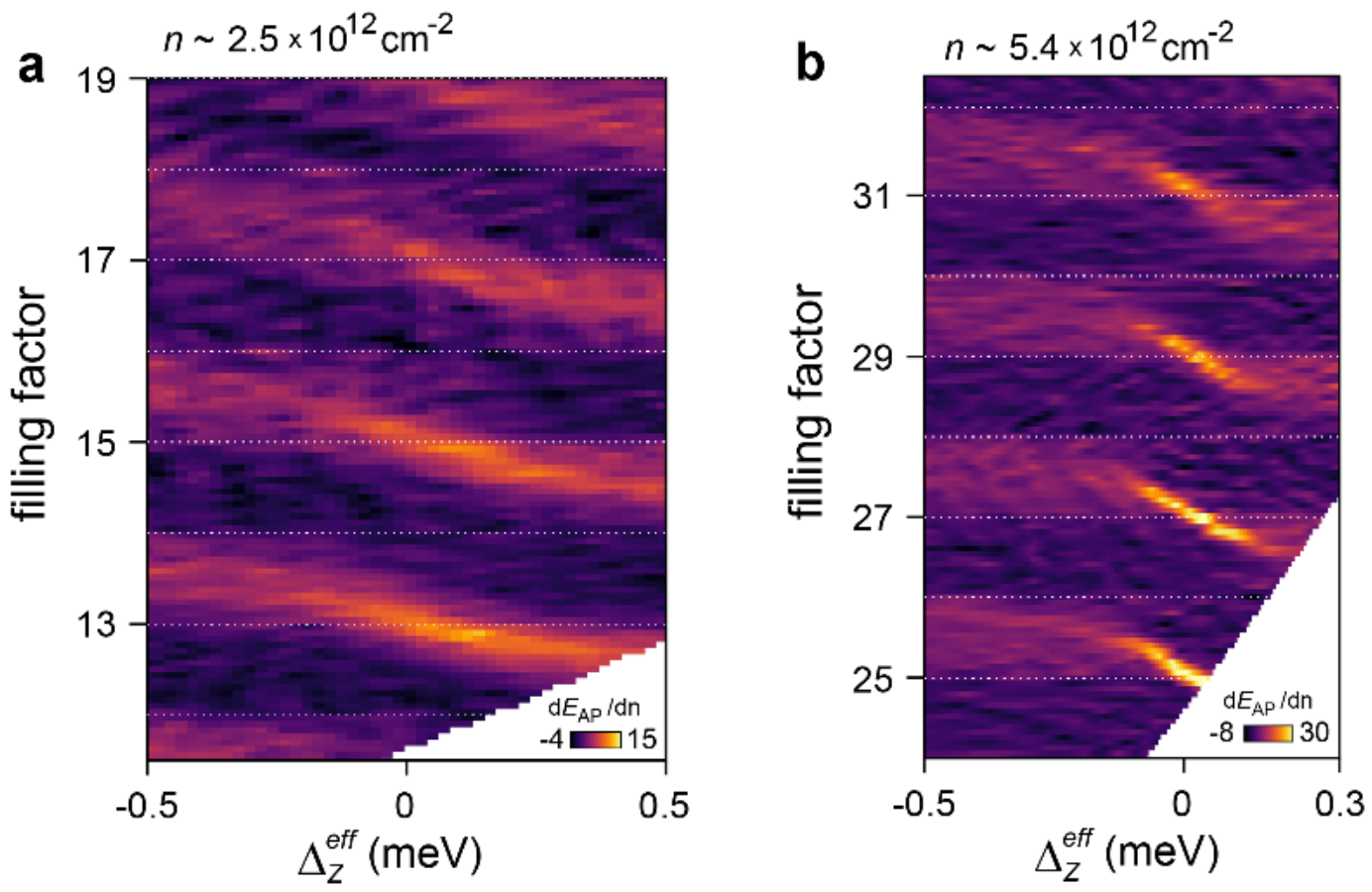

**Extended Data Fig. 5 | Landau level crossing features at additional crossing densities**

Map of the derivative of the fitted σ+ AP resonance energy with respect to the electron density, $dE_{AP}/dn$, shown versus $\Delta_Z^{eff}$ and total filling factor $\nu_{tot}$ near the three LL crossing densities (a) 2.5 × $10^{12}$ $cm^{-2}$ and (b) 5.4 × $10^{12}$ $cm^{-2}$.

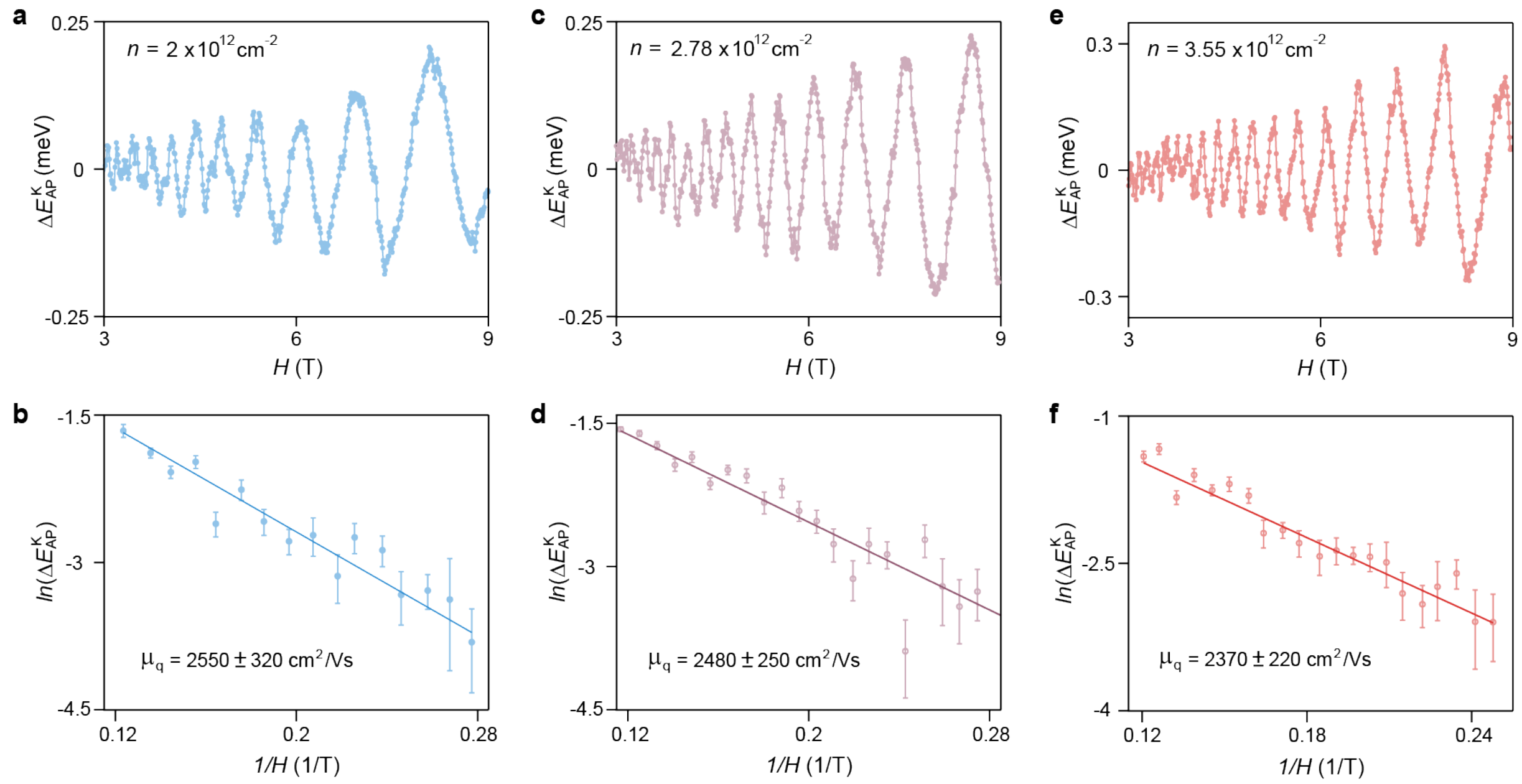

**Extended Data Fig. 6 | Quantum mobility from AP magneto-oscillations**

AP magneto-oscillations measured as a function magnetic field. Fitted AP resonances versus magnetic field at (a,b) $n = 2 \times 10^{12}$ cm$^{-2}$, (c,d) $2.78 \times 10^{12}$ cm$^{-2}$, and (e,f) $3.55 \times 10^{12}$ cm$^{-2}$, after subtraction of a smooth background. Quantum mobility was obtained from a Dingle plot analysis of the oscillation extrema.